\RequirePackage{fix-cm}
\documentclass{svjour3}	
\smartqed  
\usepackage{graphicx}
\usepackage{natbib}
\usepackage{aas_macros}

\usepackage{pbox} 

\usepackage{color} 
\definecolor{gray}{RGB}{150,150,150}

\newcommand{\degree}{$^{\circ}~$}

\begin{document}

\title{Local manifestations of cometary activity}
\journalname{Space Science Reviews}

\author{Jean-Baptiste Vincent \and
        Tony Farnham          \and
        Ekkehard K\"uhrt      \and
        Yuri Skorov           \and
		Raphael Marschall     \and
		Nilda Oklay           \and
		Ramy El-Maarry        \and
		Horst Uwe Keller}


\institute{J.-B. Vincent, E. K\"uhrt, N. Oklay, H.U. Keller \at
        DLR Institute of Planetary Research, Rutherfordstrasse, 2, 12489 Berlin, Germany\\
        \email{jean-baptiste.vincent@dlr.de}
    \and
        T. Farnham \at
        Department of Astronomy, University of Maryland, College Park, MD 20742, United States
    \and
        Y. Skorov \at
        Institut f\"ur Geophysik und extraterrestrische Physik, Technische Universit\"at Braunschweig, Mendelssohnstr. 3, D-38106 Braunschweig, Germany
    \and
        R. Marschall \at
        Physikalisches Institut, Sidlerstr. 5, University of Bern, CH-3012 Bern, Switzerland
    \and
        R. El-Maarry \at
        Birkbeck College, University of London, WC1E 7HX, London, United Kingdom
    \and
        H.U. Keller \at
        Institut fur Geophysik und extraterrestrische Physik, TU Braunschweig, D-38106 Braunschweig, Germany
}

\date{Received: 19 November 2018 / Accepted: 08 April 2019}

\maketitle

\begin{abstract}
Comets are made of volatile and refractory material and naturally experience various degrees of sublimation as they orbit around the Sun. This gas release, accompanied by dust, represents what is traditionally described as \textit{activity}. Although the basic principles are well established, most details remain elusive, especially regarding the mechanisms by which dust is detached from the surface and subsequently accelerated by the gas flows surrounding the nucleus.

During its 2 years rendez-vous with comet 67P/Churyumov-Gerasimenko, \\
ESA's Rosetta has observed cometary activity with unprecedented details, in both the inbound and outbound legs of the comet's orbit. This trove of data provides a solid ground on which new models of activity can be built.
In this chapter, we review how activity manifests at close distance from the surface, establish a nomenclature for the different types of observed features, discuss how activity is at the same time transforming and being shaped by the topography, and finally address several potential mechanisms.
\end{abstract}

\section{INTRODUCTION}
When the Giotto mission flew by comet 1P/Halley in March 1986, it completely changed our understanding of comets and their activity. Among many discoveries, the mission revealed a dark nucleus with little or no exposed ice and a non isotropic coma. While most of the surface appeared inactive, a few areas gave rise to strong collimated flows of gas and dust, commonly referred to as "jets".

In the following 30 years, five additional cometary nuclei have been imaged by space probes, confirming the picture revealed by Giotto: cometary nuclei are among the darkest objects in our Solar System, and the distribution of gas and dust density above the surface is highly non-isotropic, with ubiquitous collimated features.

\begin{figure}
  \includegraphics[width=\columnwidth]{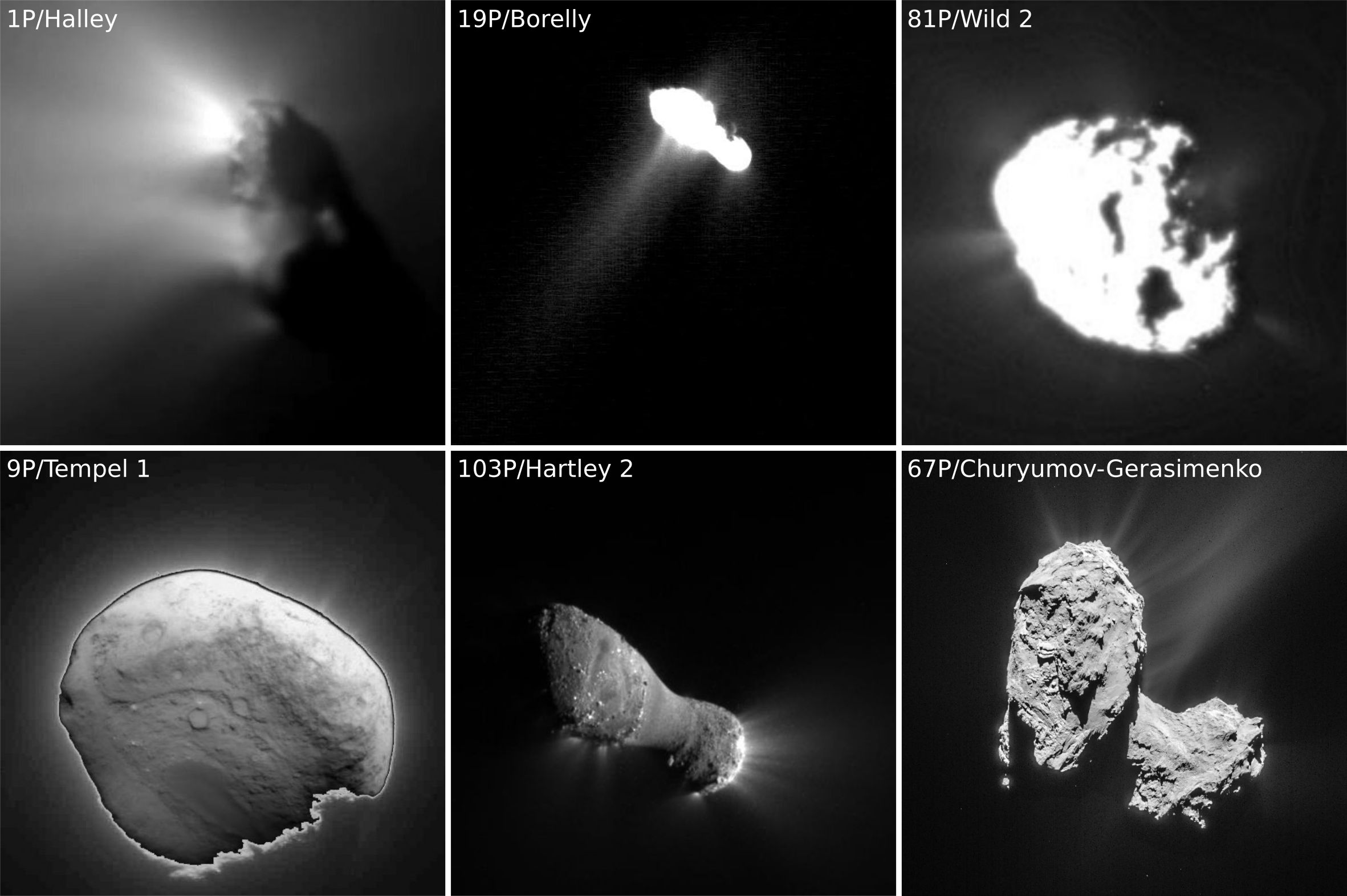}
\caption{Cometary activity observed by space missions, in chronological order from left to right, top to bottom.}
\label{fig:all_comets}
\end{figure}

While there is no doubt that coma features are collimated streams of dust and gas (for only the density contrast with the ambient coma makes them detectable in images), the cause of this collimation is challenging to establish, and has led to many publications with conflicting definitions for similar observations. One issue, to be discussed in section \ref{sec:nomenclature}, is that physical processes that are completely unrelated can give rise to the same type of morphological features. It is for instance well established that topographic variations will collimate gas flows and create regions of higher density where the dust can be more easily accelerated. This phenomenon has been detected on various objects. On the other hand, observations also suggest that some localized areas are more volatile-rich than others, with the ice being sometimes directly exposed on the surface. Consistently, this larger amount of volatile material will sustain activity for a longer time that depleted regions, thus leading to stronger gas and dust flows over those areas, and jet-like features. Figure \ref{fig:all_comets} shows how activity looked like for the 6 cometary nuclei images in situ so far.

Several attempts have been made to classify these features and connect them to processes which can then inform us about the physical and chemical properties of cometary nuclei. A comprehensive review by \cite{belton2010} proposes the following nomenclature:
\begin{itemize}
\item \textit{Type I} describes dust release dominated by the sublimation of H$_2$O through the porous mantle;
\item \textit{Type II} is controlled by the localized and persistent effusion of super-volatiles from the interior;
\item \textit{Type III} is characterized by episodic releases of super-volatiles.
\end{itemize}

In this description, \textit{Type I} jet-like features do not appear to be associated to specific morphology and are generally broader and more diffuse than other dust streams.
\textit{Type II} features, also called \textit{filaments} in the literature display a much more collimated structure and have been traced back to specific regions of cometary nuclei.
Finally, \textit{Type III} are more sporadic events, probably related to micro-outbursts or other explosive processes.\\

This description was well suited to describe observations acquired from the first flybys (comets 1P, 9P, 19P, and 81P), but does not apply to all features observed by later missions (comets 103P and 67P). In addition, this former nomenclature imposes an interpretation by associating morphological description of jet-like features with some physical processes which are still a matter of debate.

In this paper, we chose to dissociate morphology and physics: we propose a nomenclature that is purely based on observations and define terms that can remain future proof. We will show how our definitions relate to Belton's. In section \ref{sec:physics}, we will review the potential mechanisms of cometary activity, and discuss whether some processes relate more closely to one specific manifestation of activity or another.

\section{NOMENCLATURE OF COMETARY ACTIVITY}\label{sec:nomenclature}
A spacecraft orbiting a cometary nucleus will typically observe activity: release of volatile and refractory material, as producing an ambient coma and localized regions of enhanced gas/dust density.

\subsection{Ambient coma}
The ambient coma is also known as "coma background" is present in all observations and may have to be subtracted from the data in order to detect anisotropies. Is is a diffuse background of gas and dust, typically enhanced on the day side of the nucleus where sublimation is stronger.

Different volatile species will display different distributions, depending on their sublimation temperature. For instance $H_2O$ is much more abundant on the day side, while $CO_2$ expands in a more isotropic coma. Of course, this distribution is somewhat affected by local variations of composition on the nucleus' surface, but the diffuse nature of such activity makes it often difficult to associate a specific region with a given gas species abundance.
Secondary sources, e.g. ice grains ejected from the nucleus and sublimating at some distance from the surface, may introduce some anisotropy in the distribution. Example of such phenomena can be found in \cite{ahearn2011} for comet 103P/Hartley 2 or \cite{bodewits2016} for comet 67P/Churyumov-Gerasimenko

Likewise, dust also expands in a coma background, accelerated by the gas flows for a few nucleus radii until the gas density becomes too low to provide further acceleration. Beyond that limit, dust motion is mostly controlled by solar gravity and radiation pressure, and the expands into general coma, tail and trail.

A more thorough description of the coma background for dust and gas, as well as the prime mechanisms for expansion, and a summary of numerical simulations efforts on this topic can be found in the "Coma" chapter of this book (Marschall et al, 2019).

\subsection{Collimated streams}
Collimated streams have been observed around all cometary nuclei, and are always identified as narrow regions of higher density in gas/dust with respect to the ambient coma. At the distances and resolutions considered in the paper (spacecrafts flying by or orbiting cometary nuclei), these features are typically detected by stretching the contrast of images. This is different than for ground based observation which often require special image filtering in order to reveal coma structures \citep{samarasinha2014}. The absence of features in our data means that no variations in coma brightness can be detected against the noise in the background, indicating a very low level of activity, or none at all.

While they are easy to detect, naming them and understanding the physics of their formation has been a challenge ever since the first observations (either ground based or in situ). Many authors have described these collimated streams using different names: e.g. "coma structures", "jets", "jet-like features", "plumes", or "filaments".
Often the same words are used to describe collimated streams at very different scales, from spacecraft observations (a few tens of km from the surface) to ground based observations (a few tens of thousands kilometers from the nucleus).
In addition, some of these words carry a physical meaning which may not be the correct interpretation of the nature of this phenomenon. A "jet", for instance, implies that the gas/dust stream is being forced out of a small orifice. Although this is a working hypotheses, it has never been observed as such so far.
We should also be wary of misinterpretation by people with different knowledge background: "plume" describes accurately a cloud of dust/gas but also carry the general meaning of having different properties (composition, temperature) than its surrounding, and being potentially buoyant. Neither case is necessarily applicable to the case of comets. "Filament" is a better word, and already mentioned in solar physics to describe coronal streamers. It is however seldom used in the cometary literature, where "jet" (although ambiguous) has been the dominating word.\\

In this paper, we advocate the usage of neutral vocabulary and recommend the usage of "collimated streams" when describing the observations. However, we also recognize that some words are well established in the community and cannot be dismissed totally. We do strongly suggest, though, to restrain from using the single word "jet" and instead prefer the longer version: "jet-like feature". This provides a visual description of the observations that is close enough to what has been already published, while at the same time not claiming definitely that these features are produced by a jetting process.

Note that this definition is based on the morphology of collimated streams themselves and does not assume any relation between a jet-like feature and nucleus surface morphology/composition. This is an important point which will be developed in section \ref{sec:active_sources}.

\subsection{Temporality}
All missions prior to Rosetta have been single fly-bys encounters. This means that they provided a snapshot of cometary behaviour at a fixed epoch, with high resolution observations lasting only a couple of hours at best. By orbiting comet 67P for over 2 years, throughout the perihelion passage, Rosetta gave us an unprecedented view into the temporality of activity phenomena. In particular, it was possible to observe some jet-like features repeating from one rotation to the next, seemingly arising from the same area, while others were sporadic events appearing only once.

The temporal factor turns out to be extremely important when trying to infer the mechanisms leading to the release of gas and dust, and to the collimates streams we observe. In fact, a single image can often be misleading. Out of the many jet-like features visible at any time, some may last for many hours - rotating with the nucleus, switching off at night and waking up the next morning - while others may only be detected in that single frame. Yet both types of features may appear morphologicaly identical.

Some attempts have been made to include temporality in the nomenclature, e.g. \cite{belton2010, belton2013, vincent2016b}.
Unfortunately, they often mix observation and interpretation. This should be taken with caution as establishing the processes behind cometary outbursts remains a challenge. Yet, a clear distinction can be made between different types of activity based on the duration of each event. Our proposed nomenclature is described hereafter and summarized in Table \ref{tab:nomenclature}.\\

Cometary missions have observed two categories of jet-like features. While both share the same morphology and can be traced down to specific areas of the nucleus, their dynamics are quite different.\\

In the most common case, \textit{collimated streams} appear to last for many hours: they rise as soon as the local temperature enables sublimation, rotate with the nucleus, and wane some time after the local sunset. They will often be reactivated on the next morning, and repeat this cycle for many rotations. For a source more volatile than water, such features has been observed to last for several full rotations, with little or no decrease in intensity during the night \citep{feaga2007}. We will refer to these features as perennial or long-lasting.
Observations suggest that these common streams contain more dust than the surrounding coma (that's why we can detect them) but not necessarily a different speed or composition.
A particular case of perennial activity are curtain-shaped streams (i.e. collimated only in one direction) which closely follow the morning terminator and are associated to the sublimation of frost in the early local morning \citep{desanctis2015, shi2018}. While each source area will only be active for a short time, the active foot print slowly swipes a large fraction of the morning surface and is reactivated every morning, thus we consider them perennial as well.\\

The second type of jet-like features are transient, or short-lived. They arise suddenly for a few minutes, and may never be detected again from that location. These are traditionally associated to outbursts, i.e. the sudden release of a large quantity of gas and dust, driven by mechanisms that are still debated. While outbursts have usually been interpreted as akin to an explosive decompression, this interpretation is challenged by EPOXI and Rosetta data and the associated models which suggest that the transient release of material can be well explained by avalanches \citep{steckloff2016} or topographic collapses (pits, cliffs, \citet{vincent2016b}). Spacecraft observations have shown that such events are more likely to occur around perihelion and in the outbound orbit, but not always. They may arise from very different terrains and illumination conditions on the nucleus \citep{belton2013, knollenberg2015, vincent2016b, gruen2016}. Some events may also repeat from the same location, at different frequencies (from days to months, \citet{agarwal2016, oklay2018}). Transient events may lead to different morphologies for the associated dust feature. \cite{vincent2016b} distinguish three categories: \textit{Type A} are narrow, strongly collimated streams, \textit{Type B} are broad plumes, and \textit{Type C} are complex morphologies with several streams of types A and B at once. The term \textit{plume} is quite relevant here as the material associated with transient events has often different properties than the ambient coma (an order of magnitude faster, smaller grains, enriched in water ice, see papers cited above).\\

\begin{figure}
  \includegraphics[width=\columnwidth]{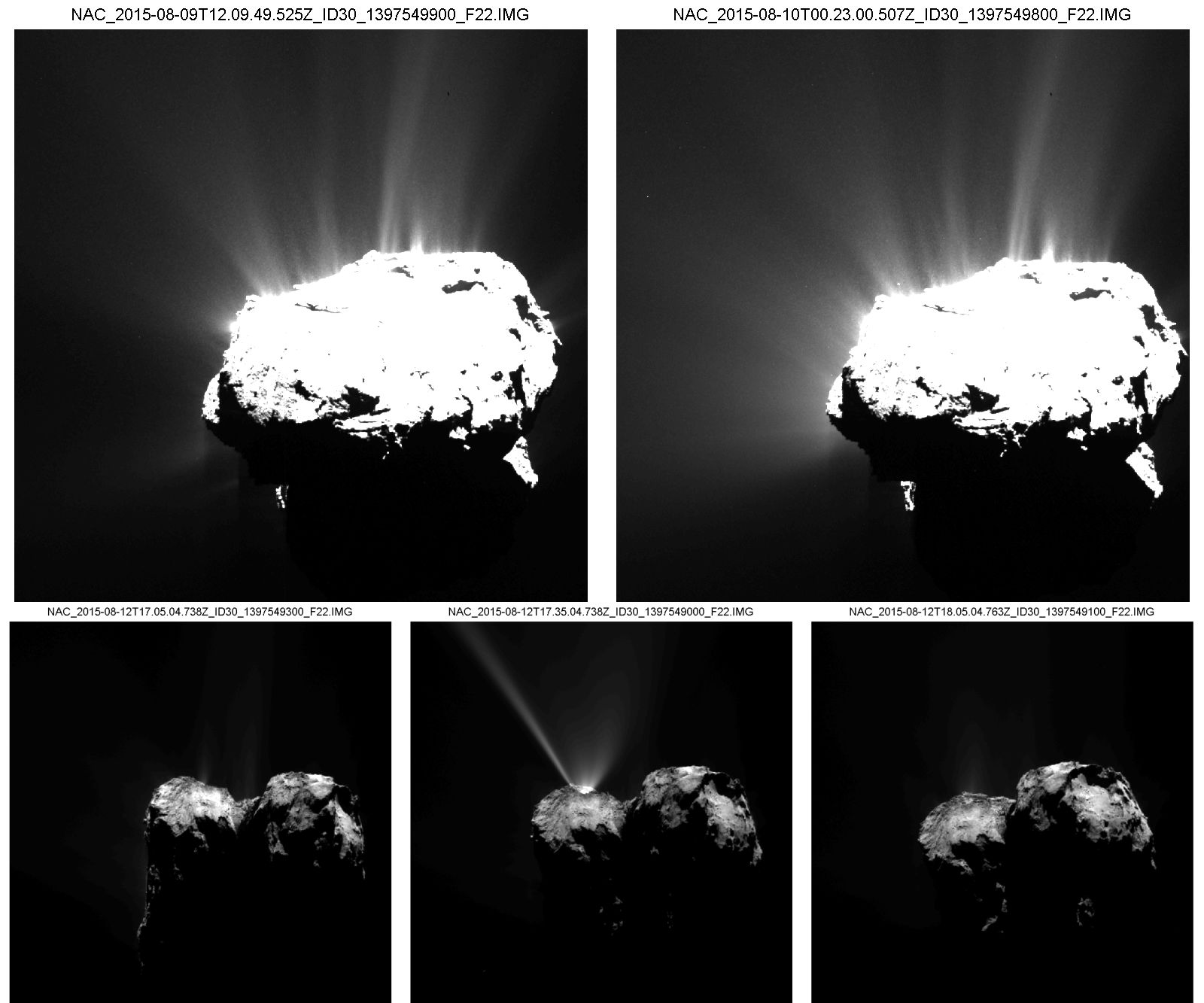}
\caption{The two main local manifestations of cometary activity, as observed by the OSIRIS Narrow Angle Camera on board Rosetta, at comet 67P. The top panels show contrast enhanced views of the comet nucleus acquired one rotation apart. One can see the exact same \textit{perennial collimated streams} consistently repeating from one comet day to the next. The bottom row shows an outburst event, in which a major \textit{transient plume appears} only in the middle frame, and is not observed 1/2h earlier or later. Adapted from \cite{vincent2016b}.}
\label{fig:jets_vs_outbursts}
\end{figure}

To summarize, we classify all local manifestations of cometary activity in two major classes: \textit{perennial collimated streams} and \textit{transient plumes}, illustrated in figure \ref{fig:jets_vs_outbursts}.
Sub-classes may be introduced when required by the observations, or when the physical process can be clearly established.\\

Going back to \cite{belton2010}'s nomenclature, our \textit{perennial collimated streams} effectively merges the types I \& II, as we seen with Rosetta that we often cannot distinguish between jet-like features related to $H_2O$ sublimation alone, or other volatiles. \textit{Transient plumes} are equivalent to Belton's type III, although once again detached from any physical interpretation.

\begin{table}
\caption{Nomenclature of local manifestations of cometary activity}
\label{tab:nomenclature}       
  \begin{tabular}{lll}
  \hline\noalign{\smallskip}
  Description &   & Classification  \\
  \noalign{\smallskip}\hline\noalign{\smallskip}
  Long-lasting events, predictable ? & $\Rightarrow$ & Perennial collimated streams    \\
                                     &              & (jet-like features) \\
  Short-lasting events, sporadic ?   & $\Rightarrow$ & Transient plumes \\
  \noalign{\smallskip}\hline
  \end{tabular}
\end{table}

\newpage
\section{HISTORY OF OBSERVATIONS}
\subsection{General introduction}
Consideration of studies of cometary activity in a historical context shows a progression in which improvements in technology and data acquisition have built upon previous knowledge to further our understanding of the physical processes at work on cometary nuclei.  It has been known from many decades of ground-based observations that active areas on the nucleus can produce features in the coma.  Linear filaments, corkscrews, arcs, and spirals are detected in images of both dust and gases, with the different morphologies reflecting the comet's activity, the dynamics of the nucleus and the viewing geometry (Farnham 2009 and references therein).  These features are often used to infer the nucleus' physical properties and to constrain characteristics of the comet's emission (e.g., features extending thousands of kilometers from the nucleus indicate that the material must be highly collimated).  Unfortunately, the large scales of ground-based observations shed little light on the interaction between the nucleus and coma or about the mechanisms producing the observed features.
With the advent of spacecraft missions that could obtain images of the resolved nucleus, it became possible to start investigating the nature of the activity itself.  Although most missions were simple flybys that captured only "snapshots" of the nucleus' behavior, each successive visit revealed additional details about the physical characteristics of coma structures.  From the very first images of 1P/Halley, through the long-term observations of 67P/Churyumov-Gerasimenko, individual comets show different types and scales of activity changing on a variety of timescales, and the aggregate of these measurements has provided the necessary background to better understand the mechanisms that drive cometary activity.  We outline here a brief history of the different comet missions, with a summary of some of the highlights of each.

\begin{table}
\caption{Spacecraft missions to comets}
\label{tab:missions}       
  \begin{tabular}{llll}
  \hline\noalign{\smallskip}
  Mission & Target & Date & Close approach distance  \\
  \noalign{\smallskip}\hline\noalign{\smallskip}
Vega 1              & 1P/Halley      &  6 Mar 1986 & 8889 km\\
Vega 2              & 1P/Halley      &  9 Mar 1986 & 8030 km\\
Giotto              & 1P/Halley      & 14 Mar 1986 & 596 km\\
Deep Space 1        & 19P/Borrelly   & 22 Sep 2001 & 2171 km\\
Stardust            & 81P/Wild 2     &  2 Jan 2004 & 237 km\\
Deep Impact         & 9P/Tempel 1    &  4 Jul 2005 & 500 km\\
Deep Impact (EPOXI) & 103P/Hartley 2 &  4 Nov 2010 & 694 km\\
Stardust NExT       & 9P/Tempel 1    & 15 Feb 2011 & 182 km\\
Rosetta             & 67P/C-G        & 2014-2016   & Orbital\\
  \noalign{\smallskip}\hline
  \end{tabular}
\end{table}

\subsection{1P/Halley flybys (Vega 1 \& 2 and Giotto spacecraft)}
1P/Halley, the only cometary mission target to date that is not a Jupiter Family Comet, was the first comet to have its nucleus resolved in spacecraft images \citep{sagdeev1987}. An important conclusion from these observations was the confirmation that the nucleus did indeed exhibit isolated active areas on its surface that produce features in the coma \citep{keller1987, larson1987}.  Unlike objects visited later, however, Halley's inner coma was so dense that it masked the details of the nucleus' surface.  This complicated any detailed analyses involving the interaction between the surface and coma and the processes that generated the activity.  A number of attempts were made to use the spacecraft images to extract information about the locations and other properties of the active areas on the surface (e.g., \cite{celnik1987}), but ambiguities in the spacecraft viewing geometry, nucleus shape and complex rotation state (e.g., \cite{belton1991, stooke1991}) made it difficult to study the coma-nucleus interactions.  There is a growing interest in revisiting the Halley spacecraft observations, using modern image processing and analysis techniques, to maximize the return from these missions and provide more up-to-date comparisons to the more recently studied comets.

\subsection{19P/Borelly  (DS1)}
Because DS1 was a technology demonstration flight redirected to comet Borrelly after the end of its primary mission, it obtained only a few high resolution images near close approach.  However, this limited dataset represented the best images of a comet's nucleus available at the time and showing a variety of different coma features, including a highly-collimated linear stream, a broad fan, and a small rapidly changing loop feature \citep{soderblom2002}.  Although the collimated streams exhibited little detail, they provided important insight that foreshadowed what was to come in future missions.

During its approach, DS1 tracked the strong, highly collimated jet-like feature for some time.  This stream was unambiguously connected to a long-term coma structure that had been detected in ground-based observations for decades, and whose constant orientation in inertial space indicates that it is aligned with the nucleus' spin axis \citep{farnahm2002}.  DS1 close approach data showed that this feature arises from a central basin at the waist of the elongated nucleus, but the limited spatial resolution of the comet's surface make it impossible to identify the source with any specific geologic structures on the surface.  Even so, this was the first measurement of a common feature that can be traced from its origin on the nucleus to $>$ 10 000 km from the nucleus, and it confirmed that features seen in ground-based observations do indeed reflect the comet's surface activity (though the comet's rotation and observing geometry can obscure the connections).

DS1 also detected the first rapid changes in a comet's innermost coma, recording a "loop" structure at the terminator where dust produced from an unilluminated portion of the surface is emerging into sunlight \citep{boice2002}.  The appearance of this structure changed on timescales of less than a minute, and it was believed that the rapid variability is related to the shutdown in activity after the source had recently entered darkness.  Again, the restricted dataset limits any detailed analysis of this phenomenon, but it represents the first occasion of features changing on very short timescales.

\subsection{81P/Wild 2  (Stardust)}
The coma of Comet Wild 2 around perihelion is dominated by a persistent, large-scale sunward fan that is centered on the nucleus' spin axis \citep{schulz2003, farnham2005a}. Unfortunately, the Stardust flyby occurred late in the apparition (98 days after perihelion), after this feature had disappeared due to the changing seasonal variations.  Thus, not only was the primary fan absent during the encounter, but the surface of the nucleus where it originated was unilluminated and could not be studied.  Stardust did record a number of small, isolated jets around close approach, however, and \cite{sekanina2004} performed an analysis of these structures.  Their results showed the features were widely distributed around the surface, with no obvious correlations in their source regions.  It should be noted that this work only projected the jet-like features back to their intersection with the surface of the triaxial ellipsoid that was fit to the nucleus, and a more complete analysis that investigates the source regions with respect to the actual shape model of the nucleus might reveal more meaningful information about specific correlations to the topography at the origins of the activity.

\cite{sekanina2004} also found that several collimated streams arose from the night side of the nucleus, confirming that activity can persist for some time after the site is no longer illuminated.  Given the locations of these features and Wild 2's 13.5-hr rotation period \citep{farnham2010} these sources may have remained active for hours after local sunset.  Unlike in comet Borrelly, the jet-like features displayed no temporal variations that would suggest the activity might be shutting down.

\subsection{9P/Tempel 1  (Deep Impact)}
Comet 9P/Tempel 1 is unique in two respects: First, it was the target of a large-scale impact experiment designed to investigate the nucleus' sub-surface characteristics.  This experiment offered the unprecedented opportunity to explore the conditions that contribute to cometary activity.  Second, it was a spacecraft target on two consecutive apparitions, offering the first opportunity to detect changes in the nucleus produced by the comet's long-term activity.

As with comet Wild 2, Tempel 1 is known to exhibit a persistent sunward fan in its coma during the pre-perihelion portion of its orbit.  Although continuum images of this feature show no short-term changes (e.g., \cite{lara2006}), CN observations resolved it into a corkscrew that oscillated on timescales consistent with the comet's 41-hr rotation period \citep{farnham2005b, schleicher2006, boehnhardt2007}.
Both observations indicate that the feature was produced by an active area located near the nucleus' south pole.  This primary fan had faded in ground-based observations by the time of the Deep Impact (DI) flyby (1 day before perihelion), the spacecraft recorded a diffuse jet-like feature arising from the south pole that likely represents residual activity from the primary fan. This stream is too diffuse to be visible against the illuminated portions of the nucleus, limiting any direct connection to surface features that might indicate its origin.
However, projecting the linear structure seen beyond the limb back toward the surface suggests it is associated with the large smooth flow that encompasses the nucleus' south pole \citep{farnham2007, vincent2010a}.

The connection between the activity and the smooth terrain was confirmed by other features seen in the DI images.  Additional polar activity was observed on the nightside beyond the terminator, where dust is seen in a string of connected sources against the dark surface.  Fortuitously, after the impact experiment, this region became illuminated by sunlight reflected off the rising ejecta cloud, and the indirect lighting revealed another, previously undetected, smooth patch on the surface, with the string of dust features aligned along its edge \citep{farnham2007, thomas2007}.
This was the first direct observation of cometary activity arising from a scarp or vertical surface, a phenomenon that has been observed on many occasions since.  The relationship between active areas and vertical faces provides a natural explanation for the persistence of cometary activity, in that a scarp can erode backward, continuously revealing fresh volatiles, while the vertical surface prevents insulating dust from accumulating and choking off the activity.\\

The Deep Impact IR Spectrometer obtained the first high quality measurements of volatiles in the coma, hinting at the processes that drive a comet's activity \citep{feaga2007}. 2007).  Notably, the southern polar region of Tempel 1 is dominated by $CO_2$ emission, while $H_2O$ is most prominent near the sub-solar point.  From these observations, it is clear that the gas production is heterogeneous, with $CO_2$ concentrated in the region where the highest activity levels are seen.  This suggests that the distribution of highly volatile material could play as significant a role in the activity as the illumination conditions.  Furthermore, the presence of $CO_2$ near the south pole suggests that it is the driver of the activity seen on the nightside of Tempel 1 (and by extension, on Wild 2 and other comets). Because $CO_2$ is more volatile than $H_2O$, even low levels of lingering heat can apparently continue to generate activity long after the region is in darkness.\\

DI also detected ice on the surface of Tempel 1, in the form of several small patches on the "flat top" that had just rotated into sunlight \citep{sunshine2006}.  These patches were associated with small, narrow streams whose proximity indicates they are related to the ice, but whose morphology suggests that they do not actually arise from it.  Instead, the jet-like feature origins were found to coincide with small dark features adjacent to the ice fields, and it was interpreted that those jets were the result of volatile sublimation which dragged sub-surface ice and dust into the coma \citep{farnham2007}. Some of this material then falls back to the surface in the vicinity of the jets, producing the ice patches that were observed.  It is notable that an increased concentration of $CO_2$ was detected in the coma around these jets, possibly permitting nighttime activity that would result in the ice fields (which are likely short-lived) being seen shortly after local sunrise.

Another constraint on the mechanisms that contribute to cometary activity comes from the DI impact experiment.  Using IR sequences of the ejecta and modeling of the excavation event, \citep{sunshine2007} showed that water ice exists within a meter of the surface, even at a site that revealed little evidence for activity in its pre-impact state.  This suggests that a thermal wave doesn't need to penetrate very deeply to sublimate volatiles.  Furthermore, on vertical surfaces that don't build up an insulating layer, ices are likely to be much closer to the surface, enabling the formation of isolated gas and dust streams.\\

In 2011, comet Tempel 1 was visited by the Stardust spacecraft on its extended mission (Stardust NExT; SDN), providing the first comparisons of a comet on two subsequent apparitions.  Combining the observations of the regions observed by both DI and SDN provides a means of investigating the surface evolution over the course of slightly more than one orbit. \citep{thomas2013} discuss these changes, the most prominent of which is the recession, by more than 50 m in places, of the leading edge of the south polar smooth flow.  The loss of as much as 108 kg of material from this region confirms that the scarp is a significant contributor to the activity arising from the polar region.

As far as activity, the SDN flyby occurred later in the apparition (+33 day), so the nucleus' production rates were much lower than during the DI flyby.  Even so, it is notable that no activity was detected from the sources seen in the DI observations, even though the viewing geometry was such that the coma features should have been seen if they were present.  The "flat top" region of the surface containing the ice patches was on the limb during the SDN departure, but it was also in darkness, so it is not known whether the small jets were still active at this point in the orbit. The conspicuous differences in this respect sets a tight constraint on how the diurnal and seasonal activity evolves throughout the orbit.

SDN did observe a number of isolated active areas \citep{farnham2013}, some of which arise from unilluminated regions, so it is not possible to evaluate their origins.  The most striking features, however, are a cluster of narrow jets seen on the limb near close approach.  These are traceable back to their sources on the surface, and \cite{farnham2013} showed that they all arise from a terraced structure that forms the boundary between a rough highland area and a smooth lowland region.  This further confirms that jet activity is frequently associated with vertical topography.

\subsection{103P/Hartley 2	(Deep Impact Extended Investigation)}

Little was known about comet Hartley 2 before the encounter, other than it is a member of the rare family of hyperactive comets (those that emit more water for their size than would be expected from water production models).  As with comet Tempel 1, the IR spectrometer observations show $H_2O$ and $CO_2$ emissions in the inner coma, with links to the continuum features seen in images at visible wavelengths \citep{protopapa2014}.  Unlike Tempel 1, however, the continuum seems to be exclusively associated with the $CO_2$ in the coma, with an almost anti-correlation to the $H_2O$.  This strongly indicates that hypervolatiles are driving the jet activity in Hartley 2.  There is also a thin layer of water ice seen on the morning terminator, visible in the approach images, that might indicate that water is recondensing or ice grains are falling back to the surface during the night \citep{sunshine2011}.  The rest of the surface appears to be free of ice, suggesting that the recondensed material rapidly sublimes away when in sunlight.

Hartley 2 exhibits numerous isolated active regions, with a strong concentration on the end of the smaller of the two lobes of the nucleus.  A second concentration is located along the evening terminator on the side of the larger lobe, and a number of more isolated jet-like features are distributed around the rest of the nucleus, including some that arise from the night side \citep{farnham2011a}.  The observed streams range from narrow, highly collimated structures to broader, diffuse fans.  Many of the jets can be traced back to their origins at scarps (including the one at the edge of the nucleus' smooth neck region), rimless depressions, and dark features that may be holes or vents \citep{brucksyal2013}.  Modeling of the activity during approach and departure indicates that the active areas seen at close approach are the main ones on the nucleus.  The activity from these sources increases and decreases with rotation/insolation, but doesn't tend to turn off completely.  Furthermore, there is little evidence that there are large sources that are inactive at the time of close approach but turn on under different illumination conditions \citep{farnham2012}.\\

In addition to the ice seen on the surface, the DI IR spectrometer also identified water ice in the coma of Hartley 2 \citep{protopapa2014}.  Much of this ice is in the form of micron-sized particles located in the jets emanating from the small-lobe.  Their correlation with the $CO_2$ emission indicates that the icy grains are being dragged from the nucleus' subsurface by the $CO_2$.  The grains sublime in the coma, producing a secondary, extended source of water that explains the comet's perceived hyperactivity.

A second population of grains, never seen in previous comet encounters, was also discovered around the nucleus of Hartley 2.  These large, individual entities are 10s of centimeters in size, and likely consist of fluffy aggregates of mixed dust and icy particles \citep{kelley2013}.  Although \cite{hermalyn2013} mapped the positions and motions of a selection of these grains, their dynamics are not well understood.  Some show no motion, and will likely return to the nucleus, others are exceeding the escape velocity but on paths that are not radial, while others are clearly escaping from the nucleus.  Because of the oddities of the particles' motions, there is no information about the origin of these grains, or whether the aggregates came directly from the nucleus or formed in the coma.\\

A final phenomenon of interest observed by the DI IR spectrometer was a water feature arising from the smooth neck region.  What makes this feature unusual is that it has no continuum associated with it.  The smooth neck has been interpreted as a gravitational low, where dust fallback has filled in the depressions, which may be producing an insulating layer that chokes off emission from the underlying surface.  Presumably, part of the fallback includes icy grains and these are likely subliming away from the surface to produce the water feature above the neck.  The fact that the escaping water does not entrain any dust grains should provide information about the mechanisms by which dust features are produced (e.g., gas arising directly from the surface may not provide a means of lifting or accelerating the dust grains.)

\subsection{67P/Churyumov-Gerasimenko}
Comet 67P/Churyumov-Gerasimenko has been observed by Rosetta from end of March 2014 to end of September 2016, from 4.3 AU inbound to 3.8 AU outbound. In this period, the distance spacecraft-nucleus varied from $5.10^6~km$ down to the surface. As this paper deals with local manifestations of activity, we will focus on the epoch August 2014 - September 2016, during which the spacecraft orbited mostly between 10~km and 300~km from the nucleus (with the exception of two far away excursions in the sunward and anti-sunward directions).

Activity is a continuous process on comets, and its intensity is strongly driven by the amount of input energy (heliocentric distance). From Rosetta observations, we can roughly define four main epochs showing significant variations in activity pattern: wake-up, equinox and perihelion, post-perihelion.

\subsubsection{Wake-up}

The beginning of the wake-up phase is not known, as the sensitivity of distant observations is limited. OSIRIS/Rosetta observations show that the nucleus was already faintly active in our first observations (4.3 AU), displaying a brightness profile slightly broader than star-like, a signature of the developing coma \citep{tubiana2015}. We do not know if this comet is active at larger heliocentric distance, but it can be investigated by a sample return mission such as CAESAR, which will approach the comet at its aphelion.

From onset of activity to arrival at the comet in August 2014, Rosetta observed the gradual development of the ambient coma, and started to resolve collimated streams at various scales. Among these features we detected several streams which appear to originate from the bright, ice-rich concavity in between the two lobes of the nucleus. These jet-like features are interpreted as the combination of enhanced volatile content in this specific area and a focusing effect by the topography \citep{lara2015, lin2015}.

It is important to note that the wake-up of activity was not only a slow and quiet process. \cite{tubiana2015} reported an outburst at the end of April 2014, which increased the brightness by 65\%. Unfortunately, the nucleus was unresolved in OSIRIS data at that time, and we do not know how the surface was modified by this event.

After Rosetta arrived at 67P on 06 August 2014, the spacecraft remained about 6 months at distances lesser than 100 km from the nucleus. Thus Rosetta was deeply into the coma, in a prime location to observe the developing activity close to the surface. \cite{vincent2015b} report on this evolution from arrival (August 2014, 3.6 AU) until equinox (May 2015, 1.6 AU). They observed that the footprints of jet-like features appear to follow the subsolar latitude, and originate from an ever larger area of the nucleus as the heliocentric distance decreases, in agreement with prediction from thermal models such as \cite{keller2015b}. They also note that activity seems to switch on and off with the Sun rising and setting over a given region. This suggests that such activity is driven by the diurnal heat wave sublimating water ice, and must originate from the surface or the very shallow subsurface.

Three-dimensional inversion of jet-like features to find their footprints show a correlation with topographically rougher areas, most notably taluses of receding cliffs and large pits \citep{vincent2015a, vincent2015b}, as it had been observed previously on other comets (e.g. 9P/Tempel 1, \cite{farnham2013}). One should note, however, that the link between morphology and jet-like features is very complex. As described by \cite{crifo2005} and many later authors, collimated streams of gas and dust may seem to arise from well defined location which are not intrinsically different from any other place. The collimation is mostly due to interactions of the flow with the topography, which explains why one is more likely to observe jets from cliff and pits rather than smooth, flat terrains. A more recent example including numerical simulations of this effect can be found in \cite{shi2018}.

It is nonetheless important to stress that the footprints of most jet-like features observed on 67P can be related to the remains of collapsed cliffs, where regressive erosion exposes icy layer previously insulated by a layer of dessicated material \citep{vincent2015b, oklay2016a}. It seems that both topography and availability of volatile material play an important role in forming the observed features.

\subsubsection{Equinox and perihelion}

The equinox crossing in May 2015 triggered a increase in cometary activity which cannot be solely explained by the reduced heliocentric distance. Due to the peculiar shape of the nucleus, and the high obliquity (52 \degree), 67P experiences strong seasonal effects. One important consequence is that the southern hemisphere remains in quasi polar night 80\% of the orbit, being illuminated only for about 1 year from one equinox to the next \citep{keller2015b}. This means that a large amount of ice can build up and be preserved on this dark side until the inbound equinox, when it is suddenly illuminated for the first time in several years \citep{choukroun2015}. This translates into an increase in the production rate \citep{hansen2016} and in the number of jet-like features observed from Spring 2015 onward \citep{lai2017}.

In addition to the increase in number of observed jet-like features, we also note that activity start to persist for some time beyond the evening terminator. \cite{shi2016} have shown that this time lag can be used to derive a thickness of the dust layer covering the buried volatile material. It typically corresponds to a few mm.

This epoch is also when Rosetta started to observe significant changes on the surface (transient pitted patterns, regressive erosion, boulder movement, fracture extension, see \citet{elmaarry2017}).

This increase of activity can also be traced by the numerous transient event detected after July 2015 onward. Despite a quasi constant monitoring, no outbursts were detected after the distant one caught in April 2014 (see previous subsection). \cite{vincent2016b} show that many major events were observed (roughly one every other comet rotation), each releasing several tens of tons of material in a few minutes. We found that outbursts occur about every 2.4 nucleus rotations. They are comparable to what had been observed on previous comets. Like for perennial collimated streams, transient plume also appear to originate from areas close to the subsolar latitude, although not always. The spatial distribution of their footprints on the nucleus correlates well with morphological region boundaries, especially areas marked by steep scarps or cliffs. In some cases, the plumes could be linked to the sudden collapse of a cliff, but many other events remained to be explained.

\subsubsection{Post-perihelion}
Further work on outbursts (e.g. \cite{lin2016, agarwal2016, oklay2018}) show that their frequency decreased after perihelion, but did not stop and transient events could be observed until the end of the mission, more than one year after perihelion.
In fact, this type of activity turns out to be extremely asymmetric: Rosetta recorded one single event in the inbound phase down to 1.3 AU, and several hundred from one month before perihelion to the end of the mission (outbound arc, 3.8 AU). This hints at outbursts being driven not by daily processes, but rather by the seasonal heat wave reaching deeper regions of the nucleus, and perhaps more volatile ices.

This asymmetry can also be found in the overall production rate \citep{hansen2016} and the number of jets seen by the cameras on board Rosetta \citep{lai2017, kramer2017}. It is compatible with our understanding of the thermal processes involved \citep{keller2015b}.

\section{ACTIVE SOURCES AND MORPHOLOGICAL CHANGES}\label{sec:active_sources}
\subsection{Active sources}
If the coma present anisotropies, and activity manifests itself as collimated streams of gas and dust, it is logical to assume an anisotropic distribution of sources on the nucleus. That is, areas where conditions are more favourable to the release of material, e.g. a local increase in volatile material, fine dust more easily liftable, or concavities leading to a heat trap. But such active regions are more elusive than it seems, and have been debated ever since the first Giotto images of comet 1P/Halley in 1984. Indeed, this data revealed a relatively homogeneous nucleus surface with almost no ice, although many collimated streams were detected in the coma \citep{keller1987}. This led the community wonder whether we actually need active and inactive areas to account for such activity (See \cite{crifo1997} and \cite{crifo2002} for an exhaustive discussion). The current consensus is that active regions do exist but the topography is a major player in focusing the gas/dust flows.

Analytical examples and numerical simulations show that it is possible to create a jet-like feature from a flat surface if the ice distribution is not homogeneous \citep{knollenberg1996}, but observations show that this situation is seldom encountered on cometary nuclei. To understand the link between activity and surface morphology, we typically  measure the position of the same jet-like features across several images, and use this information to perform a tri-dimensional stereo reconstruction of the stream geometry. Then one can calculate the intersection between the jet and the surface, keeping in mind that jets give only a faint signal and are mostly seen in limb pointing observations which may hide first few meters (or tens of meters) of the stream if the source lies beyond the local horizon. Inversion techniques like these, from spacecraft observations at close distance, are described in details in \cite{farnham2007, farnham2013, vincent2015b, shi2016} and references therein. We find that jets are more typically linked to rough terrains, where the topography present many small cavities that are very efficient at collimated surrounding streams of gas.

On comet 67P, the rendez-vous nature of the Rosetta mission allowed the teams to investigate further the potential source locations inferred from 3d-inversion of jets. We found that many features appear to arise from cliffs, with the topography naturally collimating flows of gas and dust created in the vicinity \citep{vincent2015b}. It is also important to note that the cliffs almost always display a talus enriched in water ice, thus providing a volatile input, and also indicating a recent collapse - signature of ongoing erosion \citep{oklay2016a, fornasier2017}. Transient events associated to such collapse have also been observed on 67P \citep{vincent2016b, gruen2016, pajola2017}. Other comets show similar link between activity and morphology \citep{farnham2007, farnham2013}, and also with landslides in areas covered with fine dust like on comet 103P \citep{steckloff2016}.

We also have evidence for jet-like features arising from smoother terrains \cite{shi2016, shi2018} although modeling shows that one still need some topographic variation to support the collimation \citep{shi2018}.

Overall, it seems like perennial streams preferentially arise from areas where the topography can collimate the flows of gas and dust, with a preference for places where there is more volatile material available. We note that this does preclude activity to arise elsewhere, only that it is clearly where jet-like features are the most visible.

Forward modeling of cometary activity, and link to topography, and particularly the interaction of dust and gas flows in "jets", have been published by many authors. Prominent results can be found for instance in \cite{crifo2005}, \cite{kramer2016}, \cite{marschall2017}, \cite{shi2018}. Such simulations provide a strong strong support for the topography driven model with a very repetitive coma structure with each nucleus rotation, with 90\% correlation between simulation and measured intensity around the nucleus \citep{kramer2016}.

Still, topography does not explain everything. The transient nature of outbursts suggest that some areas are likely to contain reservoirs of volatile material, released in a sporadic events when the right conditions are met. This phenomenon has been well characterized by the gas related instruments on board Rosetta. For instance, For instance, the ROSINA (Rosetta Spectrometer for Ion and Neutral Analysis) analysis has revealed that there is a strong correlation [29 out of 36] between CO2/H2O ice patches and outburst locations (\citep{kramer2017}): CO2 spots are likely candidates for later outbursts and remain active many months after the first events have been observed, indicating the presence of a reservoir extending into the subsurface. Similarly, \cite{oklay2018} have reported more than 20 outbursts originating from the same location in the Imhotep region of 67P over a period of 1 year, in the outbound arc of 67P's orbit. This once again suggests that flows from this area relate to a large buried reservoir of ices.

\subsection{Morphological changes}
If topography affects activity, one may also ask about the reciprocal. As cometary material is ejected away from the nucleus, this must change the surface.

First evidence for topographic changes in relation to activity were reported by \cite{thomas2013} for comet 9P. After one orbit, they observed the retreat of a scarp with downward erosion of $> 20m$, and morphological changes of depressions in smooth terrains. Similar large scale changes were observed on comet 67P \citep{groussin2015a}. In addition, Rosetta also discovered many other types of changes, like fault extension, cliff collapse, transport of large boulders, summarized in \cite{elmaarry2017} and Chapter "Morphology" of this book.

Dust transport from activity \citep{thomas2015b, lai2017} also play a major role in blanketing and unveiling the topography.

\begin{figure}[!ht]
  \includegraphics[width=0.9\columnwidth]{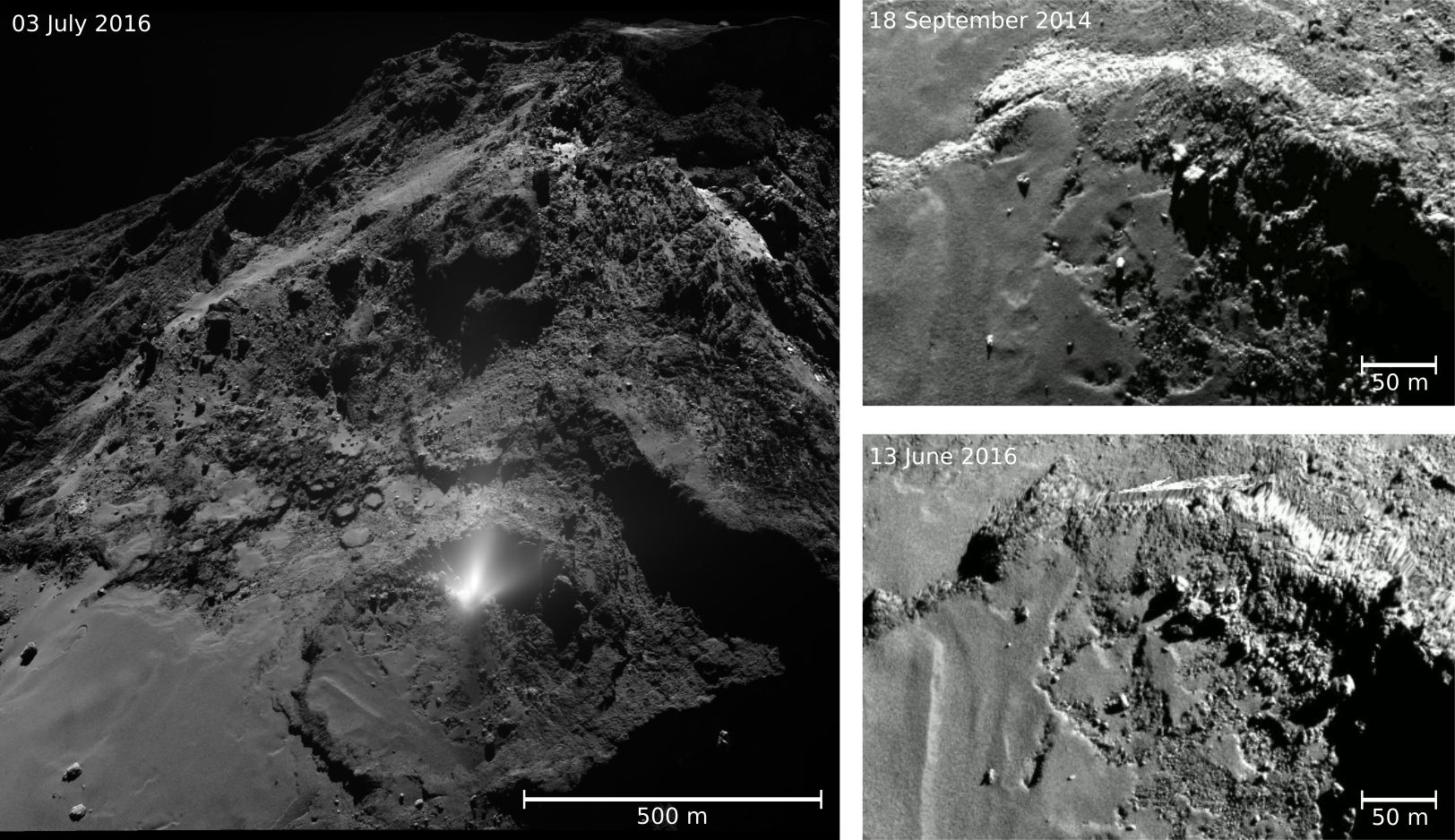}
\caption{Left: Rosetta/OSIRIS Wide Angle Camera image of the outburst plume obtained on 03 July 2016. Right: zoom in the same region before and after perihelion. Although the 03 July 2016 outburst is the only one observed at close distance, Rosetta has detected more than 20 events from this same area during the two years of the mission. The surface has changed significantly and numerous pits have been formed \cite{oklay2018, vincent2018}.}
\label{fig:outburst_imh}
\end{figure}

In addition to these large scale changes, many small scale ($< 10 m$) surface modifications have been detected and can often be associated to the sudden release of material, for instance forming small pits \citep{agarwal2016, vincent2018, oklay2018}. A prominent example from comet 67P is shown in Figure \ref{fig:outburst_imh}.

Overall, it is clear that activity affects the evolution of the surface. Its long term effects on the global shape of the nucleus starts to be understood (see for instance \cite{vincent2017}) but most of the small scale effects have only been revealed by Rosetta and remain to be explained.

\newpage
\section{NATURE OF ACTIVITY}\label{sec:physics}
A basic goal of the Rosetta mission was to learn more about the nature of cometary activity
during the long observation time. Cometary activity is the escape of gas molecules by
sublimation of ice and of dust grains from sub-$\mu$m to dm-size from the nucleus. It is widely
assumed that the dust is accelerated by the gas drag \citep{finson1968}; alternative explanations consider electrostatic forces \citep{horanyi2015}. While the dust in perennial jets seems to depart from the surface with almost zero velocity, material ejected in transient events escapes much faster and the trajectory may be controlled by the initial impulse.

The gas pressure provided by sublimating water water molecules at perihelion temperatures is in the order of 1 Pa \citep{skorov2017}. This is enough to accelerate grains up to a few decimeters in size against the weak gravity but may not be sufficient to overcome the cohesion of cometary matter and actually detach the grains of the surface. This cohesion can be due to van der Waals attraction\citep{chokshi1993} and stronger forces like sintering processes \citep{thomas1994, kossacki2015} or chemical interactions.

According to measurements \citep{attree2018b, thomas2015b, mannel2016}, the tensile strength of cometary matter reaches several Pa to kPa depending on the scale of observations.
In such a scenario of strength higher than the gas pressure one has to explain how dust can
be lifted off and how activity is maintained since an isolating dust crust may be formed. As a consequence the comet should become inactive after short time \citep{kuehrt1994}. This phenomenon of self-extinction of activity was observed in comet simulation experiments \citep{gruen1993} but obviously not at comets. This conundrum is sometimes referred to as the activity paradox \citep{blum2014}.

There are several ways out of this paradox that will be discussed in the following. When the
dust to ice volume ratio is $\ll 1$ dust grains do not touch each other and cannot form a cohesive lattice structure. They escape naturally with the subliming gas. However, according to observations \citep{rotundi2015} this is an unlikely scenario since dust seems to be the major component in comets.
An enhanced gas pressure to crack the cohesive forces would also
help. It could be achieved by considering ices with a low sublimation enthalpy. However,
such supervolatiles like $CO_2$ and $CO$ do not dominate the volatile composition \citep{hassig2015} and are probably present only in deeper layers where the heat flux from Sun is weak.
Additional heat sources in the nucleus, e.g. by chemical reactions \citep{miles2016} or phase transition from amorphous to crystalline water ice \citep{marboeuf2012} can also raise the gas pressure but there are no direct observations that prove the existence of such exothermic processes. A high pressure can also be achieved according to the pressure cooker effect \citep{kuehrt1994}. This needs basically closed pores with a good thermal contact to the surface but is difficult to believe in material with 70\% porosity \citep{paetzold2016}. Finally, \cite{vincent2012, vincent2015b, skorov2017} proposed that fractures, ubiquitous in cometary material, may act as nozzles through which the gas is accelerated. This would increase the gas velocity, thus the drag force applied to the dust.\\

An other way out of the paradox is a local weakening of the cohesive forces that can be
broken by the gas flux. Thermal fatigue is too slow \citep{delbo2014, attree2018a} to have such an effect but locally weak binding forces can be reached by other effects that were not always considered in models or lab experiments.\\

When assuming that cohesion is mainly caused by van der Waals forces \citep{chokshi1993, kuehrt1994, guettler2009, skorov2012} between spherical grains with radii R1 and R2 one
can derive the contact force \citep{chokshi1993} as
\begin{equation}\label{eq:Fc}
F_c = 3 \pi R \gamma
\end{equation}
where $\gamma$ is the surface energy per unit area and $R$ is the reduced radius
\begin{equation}\label{eq:R}
R = \frac{R_1 R_2}{R_1+R_2}
\end{equation}

According to the model of \cite{johnson1971} the cohesive strength $Y$ of a
monodisperse medium consisting of spheres of radius $R$ is given by:
\begin{equation}\label{eq:cohesive_strength}
Y = \frac{3(1-p) N_c \gamma}{4 R}
\end{equation}
where $p$ is the porosity, $N_c$ is the average number of contact points per particle. Interestingly, this strength does not depend on elastic parameters of the grains.\\

In the following the importance of several parameters that describe cometary surface layers
is discussed to figure out how dust clusters of low of cohesiveness can be formed and,
finally, be eroded by the gas.

\subsection{Material properties}
The stickiness, given by the surface energy per unit area (see Table in \cite{chokshi1993}) can vary by orders of magnitude. Water ice is a good glue ($\gamma = 0.37 J.m^{-2}$), quartz is not ($\gamma = 0.025 J.m^{-2}$). Therefore, the composition of the surface layer is of importance. However, at low temperatures, typical for cometary ices, the stickiness of ice and dust are more similar than at higher ones as they occur on Earth \citep{gundlach2015}.

\subsection{Dust/ice ratio}
The dust/ice \textit{mass ratio} $\chi_m$ derived for comet 67 P is not completely understood but \cite{rotundi2015} derived a value in the range 2 to 6. The dust/ice \textit{volume ratio} $\chi_v$, that is more important for the geometric structure (and thus for $N_c$) if we assume that small and large dust grains have the same porosity, results by multiplying $\chi_m$ with the density ratio of ice and dust. It is, therefore, considerably smaller (by about a factor 3) than $\chi_m$. A $\chi_v$ that is not too large (in the order of one or lower) gives a small scale mixture of dust and ice grains where the dust doesn't get the chance to intensely interlink over large distances and to form stable clusters to some depth. Thus the surface can be easily cleared from dust in the sublimation process independently on the cohesion. At the surface this ratio may be enhanced compared to the bulk by ice depletion due to sublimation process.

Deep Impact and Rosetta measurements \citep{sunshine2006, capaccioni2015} conclude that there are only small amounts of ice on the surface of comets. However, lab results \citep{yoldi2015} demonstrate that ice can be masked in IR spectra. Up to now, it is not completely understood to what extend and to what depth the superficial dust/ice ratio is higher than that of the bulk. The problem of determining the ratio of refractory to icy material is complex and often under-constrained. This leads to competing interpretations of the same data set. One can find the most recent discussions on this topic in \cite{levasseur_regourd2018}, \cite{fulle2019}, and Choukroun et al (this book).

\subsection{Porosity}
According to Equ. \ref{eq:cohesive_strength}, high porosity reduces the strength with $(1-p)$. Measurements on comet 67P \citep{paetzold2016} show a high porosity $p > 70\%$. Thus the strength is reduced to less than $30\%$ of nonporous granular matter.

\subsection{Grains sizes}
The cohesive strength $Y$ in a layer of equally sized spheres goes down with increasing
radius of the grains (Equ. \ref{eq:cohesive_strength}). Thus an arrangements of larger grains can be more effectively cracked by the gas than a structure of small grains. This was already pointed out by \cite{kuehrt1994} and discussed later in much more detail by \cite{gundlach2015}.
When assuming a surface energy of $0.025 J.m^{-2}$, valid for quartz, a porosity of 75\% and
four contacts per grain Equ. \ref{eq:cohesive_strength} provides a strength $Y \simeq 0.02/ R$ Pa. The cohesion between 1 cm grains is in this model 2 Pa, between dm-sized grains 0.2 Pa, which may be low enough to be torn off by gas pressure. For such large grains – that in terms of cometary structures will present hierarchically built clusters of grains - gravity is the limiting factor. In situ observations \citep{mottola2015, ott2017} and laboratory measurements \citep{gundlach2015} have confirmed that mm to dm grains that can be easily removed against cohesion and gravity and are common on the surface of comet 67P.

\subsection{Grain size distribution}
\begin{figure}
  \includegraphics[width=0.6\columnwidth]{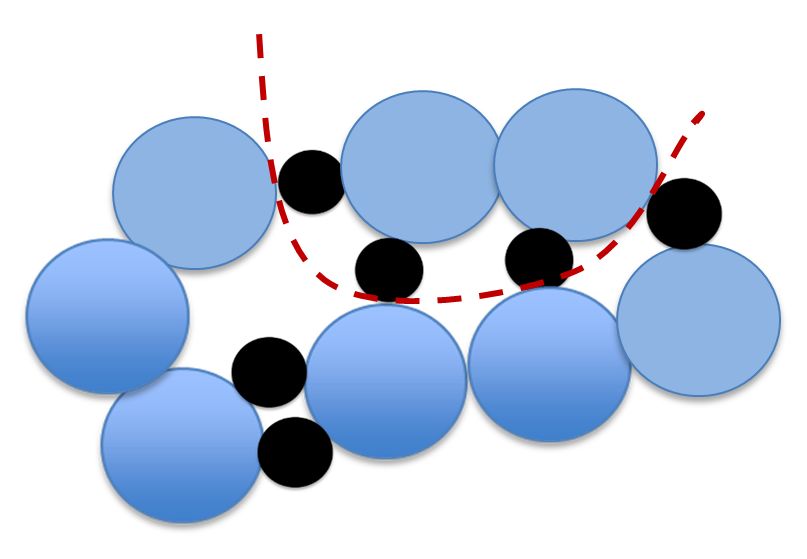}
\caption{Example arrangement of spheres of different sizes in a porous structure. The dashed red line shows a case of a cluster with a weak relative strength.}
\label{fig:grains}
\end{figure}

A further way to build clusters of low cohesion is a mixture of grains of different sizes. In such a structure (Fig. \ref{fig:grains}) with $R_2 \ll R_1$ the cohesive forces are controlled by the small (black) grains but the gas pressure can act on the cross section of the large (blue) spheres with $R_2$ . The red dashed line marks a weakly bound cluster and equations \ref{eq:Fc} and \ref{eq:R} yield:
\begin{equation}\label{eq:cohesive_strength2}
F_c = 3 \pi \gamma R_2
\end{equation}
The drag force is given by
\begin{equation}\label{eq:drag_force}
F_d = \frac{1}{2}\rho v^2 C_d \pi R_1^2
\end{equation}
where $C_d$ is the drag coefficient, $v$ the gas velocity, $\rho$ the gas density. Consequently, the ratio of drag force and cohesive binding force is $F_d/F_c \propto R_1^2 / R_2$. It is enhanced by a factor of 1000 when having a mixture of mm- with $\mu$m- or $\mu$m- with nm-particles compared to only monodisperse mm- or $\mu$m-grains, respectively.

When there is a closed line (or surface in 3D) where the local ratio $F_d/F_c$ is larger than 1 the cluster that is coincidentally formed by this surface can escape by the gas pressure.

\begin{figure}
  \includegraphics[width=0.6\columnwidth]{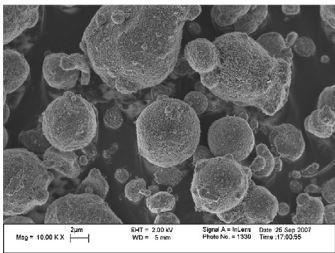}
\caption{SEM images at magnification of 10,000 of a sample with $\mu$m-sized Al particles and R972 nanoparticles for reduction of cohesiveness. Adapted from \cite{chen2010}.}
\label{fig:chen2010}
\end{figure}

An analogous effect can be reached by particles with a rough surface. It should be noted
that the described method to reduce the cohesion is known as "dry particle coating" (Fig.
\ref{fig:chen2010}) in technical applications on Earth \citep{chen2010} and is e.g. applied to improve the flow properties of printer toner. More detailed numerical simulations and lab measurements are necessary to quantify this scenario.\\

\begin{figure}
  \includegraphics[width=0.6\columnwidth]{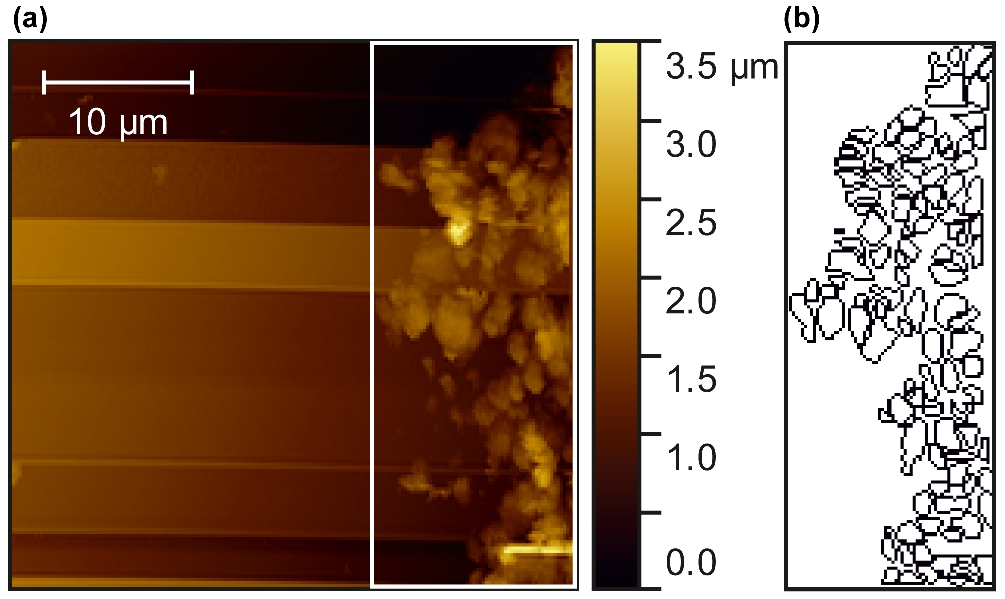}
\caption{(a) Crop of a MIDAS post-processed topographic image
(b) Corresponding area to the white frame in (a), showing all identified subunits of particle as 2D projection. Adapted from \cite{mannel2016}.}
\label{fig:mannel2016}
\end{figure}

For sure the real comet is much more complex in structure than what is shown in Fig. \ref{fig:grains} but this is valid approximation. Indeed, atomic force microscope images from the MIDAS instrument on board Rosetta revealed grains with an irregular structure (Fig. \ref{fig:mannel2016}, which can produce quite weak cohesive bounds, locally, as discussed above.

It should be mentioned that the conditions of such weak structures are mainly given near
the surface where most of the ice that clues the dust effectively together by sintering
effects has already gone. Therefore, the measured strengths of some Pa to kPa must not be in contradiction to a very low cohesive path ($<$ 1 Pa) along a given area in a surface layer.\\

\subsection{Escaping the "activity paradox"}
In summary, based on the concepts described here, we argue that the activity paradox can be addressed in two ways:\\

\begin{enumerate}
\item If the dust to ice volume ratio is $<$ 1 dust grains are isolated in the ice matrix and cannot touch each other to form a cohesive and maybe permanent dust mantle that quenches
activity. Independently on the strength of the layer the dust can be removed when the
ice is sublimating. However this scenario can only work when the dust/ice mass ratio is
small and each dust grain is immediately removed by the gas.
\item Even at a high dust/ice ration physical properties and the microstructure of the cometary surface layer result in local cohesive weaknesses that allow the gas pressure to clear the surface from dust. If irregular clusters at the surface are formed by chance that can be intrinsically stronger than the gas pressure but have weak bonds among themselves they can be lifted off even by low gas pressure. Besides such a zone of low local strength the clusters must be small enough to be removable against gravity. However, this limit is near perihelion in the meter range.
\end{enumerate}

More detailed numerical simulations and lab measurements are necessary to quantify these scenarios. In that context, future sample return missions like CAESAR (NASA New Frontier proposal) will be critical in assessing the real nature of cometary grains and their cohesion.

\section{CONCLUSION}

This chapter has reviewed what is currently known about local manifestations of cometary activity. We presented a revised nomenclature of active features, recommending the usage of descriptive but neutral terms to describe physical phenomenon we observe but do not yet understand. Throughout the paper, we use \textit{jet-like feature} when referring to perennial collimated streams, and \textit{transient plumes} for sudden events.

We listed historical in-situ observations of cometary activity over a 30 years period, from the Giotto mission in 1984 to Rosetta in 2014-216.

Finally, we dwelved into the "activity paradox", which is that current models of dust release and subsequent acceleration by gas streams formed by sublimating volatiles are not sufficient to explain our observations. While we do not have an answer to this conundrum, we do propose several ways forward, to be investigated further.


\begin{acknowledgements}
This project has received funding from the European Union’s Horizon 2020 research and innovation programme  under grant agreement No 686709 (MiARD). This work was supported by the Swiss State Secretariat for Education, Research and Innovation (SERI) under contract number 16.0008-2. The opinions expressed and arguments employed herein do not necessarily reflect the official view of the Swiss Government. This research has made use of NASA's Astrophysics Data System Bibliographic Services.
\end{acknowledgements}

\bibliographystyle{spbasic}      

\end{document}